# Nonlinear Nanophotonics for High-Dimensional Quantum States

LIAT NEMIROVSKY-LEVY[†,1,3], AMIT KAM[†,1], MEIR LEDERMAN[2], MEIR ORENSTEIN[2], UZI PEREG[2], GUY BARTAL[2] AND MORDECHAI SEGEV[1,2,3*]

[1] *Physics department, Technion – Israel Institute of Technology, Haifa 32000, Israel*
[2] *Andrew and Erna Viterbi department of Electrical & Computer Engineering, Technion – Israel Institute of Technology, Haifa 32000, Israel*
[3] *Solid state institute, Technion – Israel Institute of Technology, Haifa 32000, Israel*
*\*msegev@technion.ac.il*

**[†]These authors contributed equally to this work.**

## 1. Abstract

Quantum nanophotonics merges the precision of nanoscale light manipulation with the capabilities of quantum technologies, offering a pathway for enhanced light-matter interaction and compact realization of quantum devices. Here, we show how a recently-demonstrated nonlinear nanophotonic process can be employed to selectively create photonic high-dimensional quantum states (qudits). We utilize the nonlinearity on the surface of the nanophotonic device to dress, through the polarization of the pump field, the near-field modes carrying angular momentum and their superpositions. We then use this approach for the realization of a multilevel quantum key distribution protocol, which doubles the key rate compared to standard schemes. This idea is an important step towards experimental realizations of quantum state generation and manipulation through nonlinearity within nanophotonic platforms, and enables new capabilities for on-chip quantum devices.

## 2. Introduction

In recent years, there has been growing interest in developing quantum optics at the nanoscale, by incorporating new capabilities of photonic circuitry into nanophotonic platforms [1–5]. Quantum optics at the nanoscale holds great promise for advancing future quantum information applications, such as transmission [6], security, distributed computation [7,8] etc., as well as a promising platform for quantum computing [9,10]. Its compact size and compatibility with current on-chip technologies makes it particularly appealing [11,12]. Additionally, the tight confinement of the light in nanophotonic platforms enhances nonlinear effects such as spontaneous parametric down-conversion (SPDC) [13–15] and four-wave-mixing [16], giving rise to high flux of entangled photons. The combination of nanophotonics, nonlinear optics and quantum optics offers greater ability to control, generate, and manipulate quantum states of light. Moreover, the tight confinement of light in nanophotonics, manifested in surface modes whose intensity peaks at the surface and enable enhanced light-matter interactions, providing unique opportunities for quantum state control.

However, generating, controlling and manipulating quantum states at the nanoscale also presents new challenges, particularly in applying external control knobs at the sub-wavelength scale. Nonlinear optics can offer a powerful approach for mitigating these challenges by providing additional interaction mechanisms [17,18], which enable precise manipulation of quantum states and their interactions. These nonlinear processes have been successfully employed in various macroscopic quantum technologies, including entanglement generation, quantum frequency conversion [19,20], generating squeezed states of light [21,22], etc. While the integration of such nonlinear techniques into a nanophotonic platform is not straightforward, recent advances [23] suggest that judicious design of nanostructures can harness these processes effectively, unlocking their full potential at the nanoscale.

Nanophotonic platforms enable strong field confinement and a high Purcell factor, enhancing nonlinear parametric processes, and thus leads to higher generation rates of the desired photon states. In addition, we need to develop techniques to couple specific quantum states out, into free space, where they can be measured and applied, e.g. for the transfer of their quantum information. A recent advance, developed in our lab [24], can directly address this selective extraction of desired states from the of near-field by harnessing the optical nonlinearity inherent to metals which often comprise nanophotonic systems.

*Here, we propose a new concept for generation and manipulation of high dimensional quantum states in a nanophotonic platform, exploiting its optical nonlinearity. We present a way to generate bi-photon states carrying angular momentum by nanophotonic surface modes, and subsequently use a nonlinear interaction to project those states onto specific spin and orbital angular momentum (OAM) states of the far field photons, enabling the readout of the quantum information encoded in the near-field. This nanophotonic setting can be used as a highly compact system for encoding qudits for a multilevel quantum key distribution (QKD) protocol, offering a doubled key-rate compared to standard schemes, combined with the higher generation rates of entangled photons arising from the tight field confinement.* These platforms enable robust integration with on-chip photonic technologies, enhanced scalability, and resilience to environmental perturbations, making it uniquely suited for practical, miniaturized quantum communication devices.

### 3. Angular Momentum in Nanophotonic Systems

The total angular momentum (TAM) of light is a fundamentally conserved quantity, both classically and quantum mechanically. In specific settings such as paraxial optics, the angular momentum of the photons can be uniquely divided into spin angular momentum (SAM) and orbital angular momentum (OAM). In nanophotonic settings, which impose confinement to the sub-wavelength scale, SAM and OAM cannot be trivially separated. However, in the near-field regime, while SAM and OAM remain inseparable, photons can couple to modes characterized by their TAM, that serves as a good quantum number [25–27]. Such near-field modes carrying TAM can be entangled [28].

Traditionally, imaging and detection of near-field modes has posed a problem. This was recently resolved in a technique facilitating real-time imaging of near-field modes, relying on nonlinear polarization-dependent coupling of the nano photonic modes to far-field modes via four-wave-mixing [24]. In that scheme, the polarization of the pump field determines the polarization of the emitted photons in the far-field modes. The far-field modes can be engineered to be paraxial, such that the SAM and OAM are separable, facilitating complete reconstruction of the quantum properties originally encoded on the photons in the near-field modes. In this nonlinear imaging process, the TAM of the emitted light matches that of the nanophotonic mode. Consequently, because the TAM of the photons is the sum of the SAM and OAM, and the SAM of the measured far-field photons matches the SAM of the pump field, the measurement process enables selective measurement of the OAM of the photons, and full recovery of the TAM of the photons in the near-field modes of the nanophotonic system.

Here, as we show below, when the nanophotonic system supports multiple modes carrying different TAM per photon, combined with the nonlinear interactions described above, it enables the encoding and decoding of quantum information onto multi-level states. Thus, **the nonlinear interaction in the near field can generate high-dimensional quantum states carried by the photons that are coupled out from the nanophotonic platform**. **These high dimensional quantum states can be controlled, at will, by the pump beam.**

In the following section, we explain how the angular momentum of the emitted photons can be controlled using the nonlinear interaction.

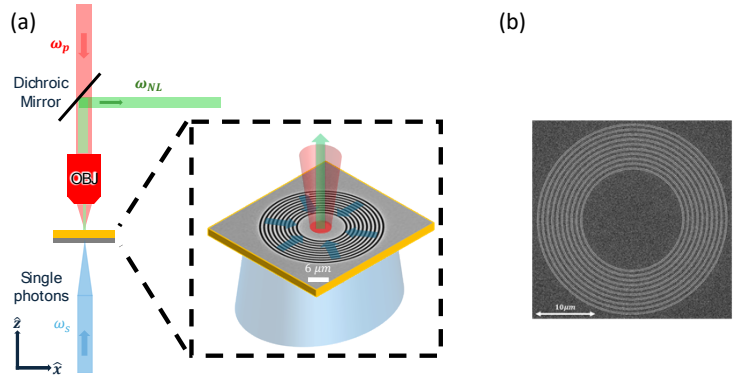

**Figure 1: (a)** Proposed setup for quantum nonlinear near-field generation of the photonic states. A photonic pump beam (red) is focused onto the sample. The plasmonic field is generated by coupling single photons (blue) into the nanophotonic modes of the sample. The nonlinear interaction yields a single photon, marked by the green arrow, coupled to far-field modes. This set-up is taken from the classical experiment in [24] **(b)** Top view SEM image of a fabricated nanophotonic device for encoding TAM multimode quantum states. The device consists of circular gratings for transforming the input entangled photon into entangled plasmonic modes. The clear central region is used for the nonlinear encoding and outcoupling process.

## 4. *Generating Qudits in the Nanophotonic System Using Linear and Nonlinear Coupling*

Qudits - quantum states with more than two levels - represent a promising avenue for quantum information processing [29]. Due to their nature, qudits allow encoding and processing of more information per mode compared to qubits [30–33], utilizing entanglement more efficiently, and enabling more compact systems. This is relevant mostly for quantum communications and teleportation of high-dimensional states [34–36], or high-dimensional quantum error correcting codes [37,38]. Moreover, using qudits for quantum computing algorithms can reduce the number of gates required to implement multi-level operations [39,40].

The generation of qudits in nanophotonic systems typically relies on advanced control over multiple photonic modes, such as angular momentum states [41] or quantum states encoded in a frequency comb [42], which offer avenues for encoding higher-dimensional information.

In our system, the qudit state generation is initiated by coupling a free space photon, in a given polarization mode, to the nanophotonic device, as shown in Fig. 1(a). The device comprises a thick gold layer that supports surface plasmon polariton (SPP) waves at the gold-air interface. The coupler is a circular slit engraved in the gold layer (concentric circular gratings, shown in Fig. 1(b), assists in enhancing the coupling coefficient). Due to the cylindrical symmetry, the coupler transforms each SAM (circular polarization) component of the input photonic state into a transverse magnetic (TM) Bessel-like plasmonic mode, characterized by its TAM, whose value is equal to the value of the original SAM of the impinging photon. The electrical field of the near-field photon (in the plasmonic mode) for right-handed circularly polarized input photon is:

$$\hat{E}_{spp}(r) = \frac{\hat{a}^\dagger_{spp}}{2} \begin{cases} k_z k_r \big(J_2(k_r r)e^{i2\varphi} - J_0(k_r r)\big)e^{-k_z \cdot z - i\omega t} \\ -i k_z k_r \big(J_2(k_r r)e^{i2\varphi} + J_0(k_r r)\big)e^{-k_z \cdot z - i\omega t} \\ 2k_r^2 J_1(k_r r)e^{-k_z \cdot z - i\omega t} \end{cases} \quad (1)$$

Where $\hat{E}_{spp}(r)$ is the electric field of the surface plasmon polariton (SPP) mode in the near-field, $\hat{a}^\dagger_{spp}$ is the creation operator for the plasmonic mode, $k_z$ is the wave vector component along the propagation direction, $k_r$ is the radial wave vector component, $J_n(k_r r)$ represents the Bessel function of the first kind of order $n$, and $\omega$ is the angular frequency of the field and $\varphi$ is the beam phase along the propagation.

Subsequently, we initiate the nonlinear parametric process that both dresses the nanophotonic mode and extract photons from the near field photonic states. Classical (coherent state) pump light field is launched vertically at the central flat surface of the device. Mediated by $\chi^{(3)}$ of the gold, the pump field interacts with the SPP photon and generates via a four-wave-mixing process an outcome photon. The interaction Hamiltonian is:

$$\hat{H}_I(t) = \frac{\varepsilon_0}{4}\chi^{(3)} \int d^3V \, E^2_{pump} \hat{a}_{spp} \hat{a}^\dagger_{out} + h.c \quad (2)$$

Where the $\chi^{(3)}$ tensor contains 21 nonzero elements, with three independent components. This results in the following nonlinear vector modes:

$$\hat{E}_{out} = \begin{pmatrix} \hat{E}_{x,out} \\ \hat{E}_{y,out} \\ \hat{E}_{z,out} \end{pmatrix}$$

$$= \chi^{(3)} \begin{pmatrix} 3E^2_{x,pump}\hat{E}^*_{x,spp} + 2E_{x,pump}E_{y,pump}\hat{E}^*_{y,spp} + E^2_{y,pump}\hat{E}^*_{x,spp} \\ 3E^2_{y,pump}\hat{E}^*_{y,spp} + 2E_{y,pump}E_{x,pump}\hat{E}^*_{x,spp} + E^2_{x,pump}\hat{E}^*_{y,spp} \\ E^2_{y,pump}\hat{E}^*_{z,spp} + E^2_{x,pump}\hat{E}^*_{z,spp} \end{pmatrix} \hat{a}^\dagger_{out} \quad (3)$$

We now analyze the emitted photon for a circularly polarized pump field, given by $\vec{E}_{pump} = E_{0,pump}\hat{\sigma}_\pm$, where $\hat{\sigma}_\pm = \frac{1}{\sqrt{2}}(\hat{x} \pm i\hat{y})$. The electrical field of the output photon resulting from the nonlinear interaction is described as:

$$\hat{E}_{out} = \chi^{(3)}(E_{0,pump})^2 \begin{pmatrix} \hat{E}^*_{x,spp} \pm i\hat{E}^*_{y,spp} \\ \pm i(\hat{E}^*_{x,spp} \pm i\hat{E}^*_{y,spp}) \\ 0 \end{pmatrix} \hat{a}^\dagger_{out} \quad (4)$$

$$= \chi^{(3)}(E_{0,pump})^2 \hat{E}^*_{\sigma_\pm,spp} \hat{a}^\dagger_{out} \begin{pmatrix} 1 \\ \pm i \\ 0 \end{pmatrix}$$

Where $E_{\sigma_\pm,supp} \equiv \frac{E_{x,spp} \pm iE_{y,spp}}{\sqrt{2}}$. By substituting the plasmonic field from Eq. (1) into Eq. (4), the emitted single photon is:

$$\hat{E}_{out} = \begin{pmatrix} \hat{E}_{\sigma_-,out} \\ \hat{E}_{\sigma_+,out} \\ \hat{E}_{z,out} \end{pmatrix} = \chi^{(3)}(E_{0,pump})^2 E_{0,spp} \begin{pmatrix} -k_z k_r J_{n-1}(k_r r)e^{i(n-1)\varphi} \\ k_z k_r J_{n+1}(k_r r)e^{i(n+1)\varphi} \\ k_r^2 J_n(k_r r) \end{pmatrix} \hat{a}^\dagger_{out} e^{ik_z \cdot z - i\omega t} \quad (5)$$

This result clearly shows that the field components depend on the rotating in-plane field $\hat{E}^*_{\sigma_\pm,spp}$ of the single plasmon, which is determined by the classical pump field polarization $E_{0,pump}\hat{\sigma}_\pm$. That is, $\hat{E}_{out} \propto \frac{1}{\sqrt{2}}|\sigma_+ \pm \sigma_-, \ell_{n-1} \pm \ell_{n+1}\rangle$. Consequently, the nonlinear interaction facilitates the emission of a single photon whose vector components are directly influenced by the near-field characteristics, whether expressed in a Cartesian or a circular basis. This process enables precise control over the angular momentum of the emitted photon by adjusting the classical pump polarization.

Next, we consider the coupling of a biphoton state into the nanophotonic system, allowing to explore the nonlinear interaction with near-field modes. This process begins by generating an entangled photon pair from a type-II spontaneous parametric down-conversion (SPDC) crystal, which in the circular polarization basis, is described as:

$$|\psi\rangle_{AB} = \frac{1}{\sqrt{2}}(|\sigma_+\rangle_A|\sigma_-\rangle_B + |\sigma_-\rangle_A|\sigma_+\rangle_B) \quad (6)$$

Here, $|\sigma_+\rangle, |\sigma_-\rangle$ indicate right and left circular polarizations, respectively. Labels $A$ and $B$ refer to the two photons. As explained above, the polarization of the output photon is determined by the polarization of the pump in the four-wave-mixing process.

It is instructive to explain the role of the TAM and OAM of the near field plasmonic modes. An intuitive way to express the vector-field of the plasmonic mode is through the decomposition of the in-plane field components in terms of left-handed ($LH, \sigma_+$) and right-handed ($RH, \sigma_-$) rotating components. In this representation, a TAM value of $J = n\hbar$ in the near field mode is manifested in a z-component with spatial distribution of the $n^{th}$-order Bessel beam (carrying OAM of $L = n$), and the in-plane components: $\sigma_-$ component with spatial distribution of the $(n + 1)^{th}$-order Bessel beam (carrying OAM of $L = n + 1$) and

$\sigma_+$-component with the spatial distribution of the $(n-1)$th-order Bessel beam (carrying OAM of $L = n - 1$). It is easy to see that $J = L + \sigma$ is the same for all field components.

Coupling the biphoton states, the following nanophotonic plasmonic modes achieved on the surface:

$$|\psi\rangle_{AB} \rightarrow |1_{J_1}\rangle_A |1_{J_{-1}}\rangle_B \qquad (7)$$

Where $|\psi\rangle$ is the initial biphoton state coupled into the nanophotonic system, $|n,m\rangle$ are Fock states corresponding to the rotating plasmonic mode basis which consists of Bessel modes with TAM values $+\hbar$ and $-\hbar$ for $J_1, J_{-1}$ respectively. For the derivation of the nanophotonic states, see the supplementary information.

Up to this point, we have two distinct levels stored in the nanophotonic quantum states encoded by their TAM values. Now, to generate the final qudit state, we extract the photons from the system to a measurement (or communications) channel. This is achieved through the nonlinear parametric process between the SPP modes and an intense classical field ("pump"), as described above and in [24]. The two extracted photons, now propagating to the far field, are in different orthogonal entangled states of SAM and OAM values out of two possibilities:

$$\begin{aligned} \text{for } \sigma_+ \text{ polarized pump} &\rightarrow |\sigma_+\rangle_{S_A} \otimes (|j_0\rangle_L + |j_{-2}\rangle_L) \\ \text{for } \sigma_- \text{ polarized pump} &\rightarrow |\sigma_-\rangle_{S_A} \otimes (|j_0\rangle_L + |j_2\rangle_L) \end{aligned} \qquad (8)$$

As shown in the Appendix, **their polarization matches that of the pump field, while their TAM value corresponds to the that of the interacting SPP mode. Thus, each of the extracted photon can now have one of four possible values** (presented in Fig. 2). For instance, if the pump field is circularly polarized, two values are retained according to the TAM of the plasmonic mode ($J_1$ and $J_{-1}$) and two values corresponding to the two possibilities of polarization handedness of the pump field. Similarly, with a linearly polarized pump, the zero TAM of the SPP modes in the alternative linear basis is preserved, offering a different set of four states. This approach allows for precise control over the generated qudit states, enabling flexible quantum information encoding.

The quantum state of the far-field photons can now be expressed on an extended basis as $|S_A\rangle \otimes |\ell\rangle$,

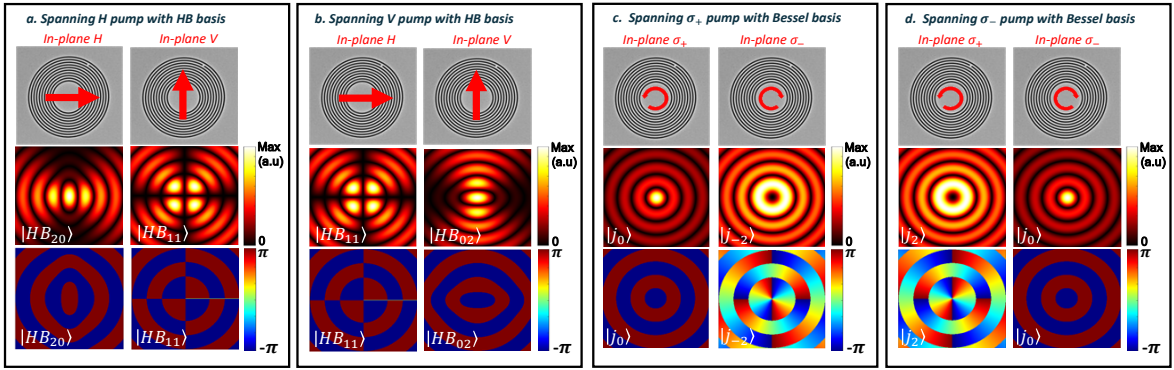

**Fig. 2**: Spatial distributions of photonic states illustrating the amplitude and phase profiles based on the polarization of the pump. The spatial patterns are represented either by Bessel beams or Hermite-Bessel (HB) functions, which can be described as superpositions of Bessel beams. When the nonlinear process involves a circularly polarized pump, the emitted photons are described by Bessel beams (as shown in Fig (c) and (d)). In contrast, a linearly polarized pump (Fig (a) and (b)) results in photon emission described by HB functions.

where $S_A$ represents its spin (polarization), and $\ell$ is the OAM DoF. The polarization is a binary DoF,

while the OAM DoF, also known as related to a topological charge $\ell$, can take any value $\ell = m\hbar$, where $m$ is integer. The spin of the emitted photon is determined solely by the pump polarization, consequently the OAM of the emitted photon is determined by the subtraction of this spin from the TAM of the original near field mode. When the emitted photon is in a paraxial mode, the SAM and OAM of the far-field photon become separable, enabling to recover the quantum information encoded on the photon by projections onto SAM and OAM states separately.

### 5. Nanophotonics-enhanced QKD

In this section, we propose the use of these qudits encoded in the far-field photons for a quantum key distribution (QKD) protocol. This allows the sender to encode each digit of the key by choosing one of the bases. A photon is then prepared in a specific polarization state corresponding to the digit value, with the selected basis defining the encoding scheme. Building on the well-known BB84 protocol [43,44], our scheme extends its framework to a higher-dimensional version by employing qudits instead of qubits. The BB84 protocol is a foundational QKD scheme designed to establish a secure communication channel between two parties, Alice and Bob, by exploiting the principles of quantum mechanics. By encoding information in qudits across two distinct bases, the proposed scheme not only generalizes the BB84 protocol but also offers the potential for enhanced key rate, as qudits carry more information per photon compared to qubits [45]. This high-dimensional approach improves the efficiency of the protocol, demonstrating the utility of the generated qudits for advanced quantum communication applications.

Our approach utilizes the nanophotonic platform presented in Fig. 1 to generate a shared key based on the SAM and the OAM DoFs of the far-field photons released from the nanophotonic platform. As we show below, incorporating multiple DoFs allows to encode quantum information on multilevel bases, paving the way to enhanced coding density compared to standard protocols relying only on the binary polarization DoF (as in BB84).

In our protocol, Alice and Bob utilize a classical channel for public communications, through which they announce their respective choices of measurement bases for their photons. Due to the conservation of TAM and the predefined quantum states generated in the near field, Alice can predict the outcome of Bob's measurement once she measures her photon — unless an eavesdropper has interfered.

We now outline the QKD protocol step by step:

a. *Randomization*

In order to distill a secret key, we need to generate two random processes, comprising the sequence values and transmission bases.

In our realization, the basis selection simply switches between a circular and linear polarization of the pump field. The data sequence is obtained from a sequence of four-dimensional qudits, which are generated by two sequential binary random selections: one is the specific polarization handedness (direction) of the pump field, as explained above, and the second is the random selection of the photon TAM value which is performed by a beam splitter (see Fig. 3). The nanophotonic system stores two SPPs with different TAM as described in eq. 7. When extracted by the pump field they pass through a 50:50 beam splitter. At the output, one of the photons is directed to the communications channel (i.e., to Bob). and the other goes to Alice's measurement module. All other output combinations of the beam splitter are not viable for the key generation. Alice measures her photon and based on the measurement outcome, she deterministically knows which photon was sent to Bob. This measurement also allows Alice to herald and synchronize Bob's measurements.

**The logical ququart states, $|0\rangle_L, |1\rangle_L,$ and $|2\rangle_L$ and $|3\rangle_L$, which are associated with the digits of our secret key**, are then encoded, as presented in the next section.

### b. *Encoding*

As a result of the process – the photons sent to Bob are four valued qudits, which carry their information in their SAM and OAM DoFs. In the circular polarization basis, the extracted bi-photon state is $|\sigma_+\rangle_{S_A}$ $\otimes (|j_0\rangle_L + |j_{-2}\rangle_L)$ for $\sigma_+$ polarized pump or $|\sigma_-\rangle_{S_A} \otimes (|j_0\rangle_L + |j_2\rangle_L)$ for $\sigma_-$ polarized pump. Let $j_n$ ($n = 0, -2, 2$) indicate a Bessel beam carrying $n$ units of angular momentum. As explained above, one photon from each pair is sent to Bob, while the other is locally measured by Alice. We assign the digits 0,1,2,3 to the following four photonic states sent to Bob by applying the circularly polarized pump field basis:

$$\begin{cases} "1" = |\sigma_+, j_0\rangle \\ "2" = |\sigma_-, j_0\rangle \\ "3" = |\sigma_-, j_2\rangle \\ "0" = |\sigma_+, j_{-2}\rangle \end{cases} \quad . \tag{9}$$

Thus far we have analyzed the case where the pump field is circularly polarized. However, to realize the BB84 protocol, it is essential to use two mutually unbiased bases. The other basis would naturally be the basis of linearly polarized pump. The biphoton state which is coupled to the nanophotonic device in this case would be

$$|\psi\rangle_{AB} = \frac{1}{\sqrt{2}}(|H\rangle_A|H\rangle_B + |V\rangle_A|V\rangle_B) \tag{10}$$

Where $|H\rangle, |V\rangle$ indicate horizontal and vertical linear polarizations, respectively, and labels $A$ and $B$ refer to the two photons.

For convenience of tracking the nonlinear process we rewrite the dual mode nanophotonic state of eq. 7 in the combination state:

$$|\psi\rangle \rightarrow \frac{1}{\sqrt{2}}(|2_{J_+}, 0_{J_-}\rangle + |0_{J_+}, 2_{J_-}\rangle) \tag{11}$$

where, $|n,m\rangle$ are Fock states corresponding to the plasmonic mode of azimuthal standing waves $J_\pm$ that are defined as $J_\pm = \frac{1}{\sqrt{2}}(J_1 \pm J_{-1})$.

With linearly polarized pump field the output photon spatial distribution can be represented by Hermite-Bessel functions which are proper superposition of the Bessel beams [46,47]. The $|2_{J_+}, 0_{J_-}\rangle$ part yields a photon at a state $|H\rangle_{S_A} \otimes |HB_{20}\rangle_L$ for $H$ polarized pump, or $|V\rangle_{S_A} \otimes |HB_{11}\rangle_L$ for $V$ polarized pump. Similarly, the $|0_{J_+}, 2_{J_-}\rangle$ part results in $|H\rangle_{S_A} \otimes |HB_{11}\rangle_L$ for $H$ polarized pump, or $|V\rangle_{S_A} \otimes |HB_{02}\rangle_L$ for $V$ polarized pump, where $|HB\rangle_{nm}$ are Hermit-Bessel beams that do not carry angular momentum. As

before, one photon in each pair is sent to Bob and the second measured by Alice. We assign the four digits for the linearly polarized pump field as:

$$\begin{cases} "1" = |H, HB_{20}\rangle \\ "2" = |V, HB_{11}\rangle \\ "3" = |H, HB_{11}\rangle \\ "0" = |V, HB_{02}\rangle \end{cases} \quad (12)$$

The knowledge of TAM $J$, the pump polarization, and the OAM collectively facilitate the generation of the secret key, according to equations (9) and (12). The different OAM states that can be measured are shown in Fig. 2.

As shown in the supplementary information, there are eight potential measurement outcomes for both Alice and Bob, and each of these outcomes represents one of the four possible digits in the secure key according to the encoding described above.

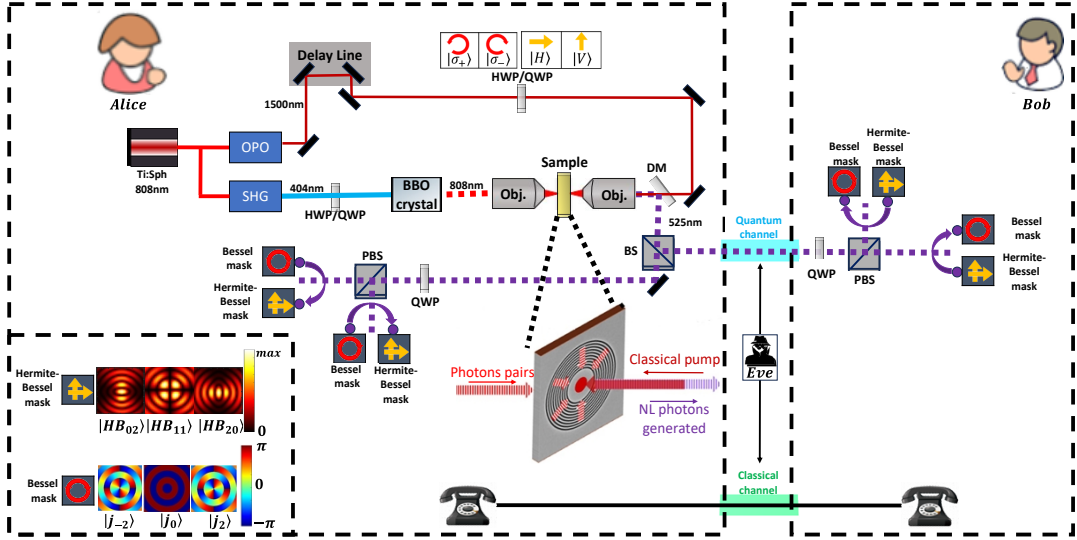

**Fig. 3. Proposed Experimental Configuration for the Nanophotonic QKD Protocol:** Alice utilizes a short pulse laser at 404 nm photons followed by an SPDC process in a BBO crystal, and an OPO to generate a 1500 nm pump. She launches the entangled biphoton state to the nanophotonic sample, giving rise to the quantum nanophotonic modes. Illumination of the sample from the opposite side with a classical pump, which is generated by OPO using the same pulsed laser source, initiates non-linear interactions within the nanophotonic field. This interaction leads to the emission of new photons, which one of them is measured by Alice and the other one is sent to Bob through a quantum channel. Bob selects a mask, which is matched either to Bessel or HB mode pattern, and detects the OAM of the photon. In addition, Bob and Alice share a classical channel for public discussion. The setup includes half-waveplate (HWP) and quarter-waveplate (QWP), beam splitter (BS) and dichroic mirror (DM) incorporated into the system.

c. *Transmission*

Bob receives the transmitted photon and randomly selects one of the two measurement bases (circular or linear polarization). If the circular polarization is selected, he transfers the photon through a quarter waveplate followed by a polarization beam-splitter and, using a proper phase mask, the orbital angular momenta of the photons are detected: $\{|j_0\rangle, |j_2\rangle\}$ in the $\sigma_-$ arm and $\{|j_{-2}\rangle, |j_0\rangle\}$ in the $\sigma_+$ arm. On the other hand, if the linear polarization basis is selected, Bob transfers the photon through a polarization beam splitter, and then, by using a matched amplitude filter, the photon with the proper Hermit Bessel order is detected: $\{|HB_{11}\rangle, |HB_{20}\rangle\}$ in the H arm and $\{|HB_{02}\rangle, |HB_{11}\rangle\}$ in the V arm (figure 3). The derivation of the possible modes achieved by the chosen mask is elaborated in the supplementary

information. The successful transmission relies on the condition that **Alice and Bob must select identical bases for both the pump and the OAM mask used by Bob**.

      d.   *Public Discussion and Information Reconciliation*

Next, Alice and Bob publicly announce their selections: Alice reveals the pump polarization type (circular or linear), and Bob reveals the basis of his measurements (either circular or linear). They discard any instances where the basis does not match. From the remaining sub-sequence, they distill the secret key. Figure 4 depicts a truth table encompassing all possible measurements, assuming that the participants have selected identical bases.

Similar to the BB84 standard protocol, Alice refrains from publicly disclosing the entire state of the transmitted photon. She only declares the pump polarization and her initial biphoton state, without revealing the OAM she measured. Bob, from his side, shares the basis of the mask he employed to measure the OAM, but does not disclose the specific OAM he measured. This approach ensures that a potential eavesdropper monitoring the public channel during the information reconciliation stage obtains only partial information.

      e.   *Error Check*

In the previous step, Alice and Bob publicly reveal and compare the choices Alice made for her pump and Bobs' basis mask. Bob can choose either one of two possible masks to measure the OAM: if he chooses a circular mask then the possible OAM states can be $\{|j_0\rangle,|j_2\rangle,|j_{-2}\rangle\}$ (see fig. 2(c-d)). If he chooses the linear basis, he can measure an OAM state from $\{|HB_{02}\rangle,|HB_{20}\rangle,|HB_{11}\rangle\}$ (see fig. 2(a-b)). A discrepancy in the spatial distribution he observes, whether it is a measurement outcome that differs from the 4 possible outcomes, or a different basis compared to the basis Alice chose for her pump at the level of the public discussion, would indicate potential interference from an eavesdropper. In addition, since the two photons emitted from the nanophotonic sample are correlated (in the circular basis if one has a TAM $J_1$ then the other photon must have TAM $J_{-1}$, and in the linear basis the two photons must have the same TAM). In any case where Bob measured TAM different than these two conditions, he knows that an error has occurred. Additionally, by measuring his own photon, Bob can deduce what OAM state Alice has measured and vice versa.

*Secure Key Generation and Privacy Amplification*

At the end of the public discussion, Alice and Bob use the remaining string to distill a shared key. Two modes are formed in the nanophotonic platform. Alice measures one mode, and the other is transmitted to Bob. Leveraging the correlation between the two modes, Alice can infer the TAM $J$ of the state of the photon transmitted to Bob, enabling the extraction of encoding and the generation of a secure key. A truth table illustrating the various potential states in which Alice and Bob can engage is presented in Figure 4.

| Alice random classical pump | $\|\sigma_-\rangle$ | $\|\sigma_-\rangle$ | $\|\sigma_+\rangle$ | $\|\sigma_+\rangle$ | $\|H\rangle$ | $\|V\rangle$ | $\|H\rangle$ | $\|V\rangle$ |
|---|---|---|---|---|---|---|---|---|
| Alice's OAM measurement | $j_0$ | $j_2$ | $j_0$ | $j_{-2}$ | $HB_{20}$ | $HB_{11}$ | $HB_{11}$ | $HB_{02}$ |
| TAM state of Alice | $J_{-1}$ | $J_1$ | $J_1$ | $J_{-1}$ | $J_+$ | $J_+$ | $J_-$ | $J_-$ |
| Alice's secret key | 3 | 2 | 0 | 1 | 1 | 2 | 3 | 0 |
| Bob's random mask | | | | | | | | |
| Bob's OAM measurement | $j_2$ | $j_0$ | $j_{-2}$ | $j_0$ | $HB_{20}$ | $HB_{11}$ | $HB_{11}$ | $HB_{02}$ |
| TAM state of Bob | $J_1$ | $J_{-1}$ | $J_{-1}$ | $J_1$ | $J_+$ | $J_+$ | $J_-$ | $J_-$ |
| Public discussion | | | | | | | | |
| Shared secret key | 3 | 2 | 0 | 1 | 1 | 2 | 3 | 0 |

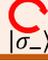

**Fig. 4. |Truth table of all possible digits in the cryptographic key using the nanophotonic QKD protocol.** Each participant, including Alice and Bob, may have multiple potential digits in their respective keys. The secret key generation entails publicly declaring their measurements. The brown frame represents the public discussion (via a classical channel) from Alice and Bob's side – Alice declares her pump polarization, and Bob declares his random measurement mask. If they measure in different bases, akin to the BB84 protocol, the procedure is halted. Conversely, if their measurements align in the same basis, both Alice and Bob can deduce the exclusive secret digit they share. The green frame represents the quantum channel, where the photon emitted from the nanophotonic sample is transmitted to Bob.

## 6. Conclusions

In summary, we presented a novel approach which employs nonlinear optics in nanophotonic platforms for the generation and manipulation of high-dimensional quantum states. By exploiting a nonlinear interaction and the properties of the total angular momentum carried by near field photons, our method enables the generation of photons with predesigned specific SAM and OAM combinations. The ability to control quantum states at the nanoscale paves the way for practical implementations of quantum information protocols, such as multilevel QKD with enhanced key rates, analyzed in the supplementary information, due to the use of qudits instead of qubits, within compact, scalable, and robust nanophotonic devices. Notably, this work demonstrates how near-field properties can be utilized for encoding quantum information and executing communication tasks.

Looking forward, incorporating such nanophotonic platforms within on-chip technologies represents a significant step toward realizing scalable quantum computing and communication systems. By combining the strengths of nonlinear optics and nanophotonics, this work paves the way for advancing future quantum technologies.

**Disclosures.** The authors declare no conflicts of interest.

**Data availability.** Data underlying the results presented in this paper are available in Supplement 1.

See Supplement 1 for supporting content.